\documentclass[letterpaper,12pt]{article}
\usepackage{tabularx} 
\usepackage{amsmath}  
\usepackage{float}
\usepackage{indentfirst}
\usepackage{graphicx} 
\usepackage[margin=1in,letterpaper]{geometry} 
\usepackage{caption}
\usepackage{cite} 
\usepackage[final]{hyperref} 
\usepackage{tensor}
\hypersetup{
	colorlinks=true,       
	linkcolor=blue,        
	citecolor=blue,        
	filecolor=magenta,     
	urlcolor=blue
}
\begin{document}

\begin{center}

\vspace{10pt}
\large{\bf{Entropy in the interior of a Kerr black hole}}

\vspace{15pt}
Xin-Yang Wang, Jie Jiang, Wen-Biao Liu

\vspace{15pt}
\small{\it Department of Physics, Beijing Normal University, Beijing 100875, China}

\vspace{30pt}
\end{center}
\begin{abstract}
{\color{red}{Christodoulou $et. \, al$}} have shown that {\color{red}{the interior volume }}of a Schwarzschild black hole grows linearly with time. {\color{red}{Subsequently}}, their conclusion has been extended to the Reissner{-}Nordstr$\ddot{\text{o}}$m {\color{red}{(RN)}} and Kerr black holes. Meanwhile, the entropy of the scalar field inside a Schwarzschild black hole has also been calculated. {\color{red}{In this paper, a general method calculating the number of quantum states of the scalar field inside the black hole}} is given, {\color{red}{which can be used in an arbitrary black hole. After introducing the two important assumptions as the blackbody radiation assumption and the quasi-static process assumption, the entropy of the scalar field inside a Kerr black hole is calculated  using the differential form, and we find that the variation of the entropy is proportional to the variation of the Bekenstein-Hawking entropy except the ending of the black hole evaporation. Similarly, we recalculate the entropy of the scalar field inside a Schwarzschild black hole and demonstrate that the entropy inside a Kerr black hole can exactly degenerate to the Schwarzschild black hole. Finally, we find that the proportionality coefficient between the entropy of the scalar field and the Bekenstein-Hawking entropy in Schwarzschild case, which is obtained using the differential form, is half of that given in the previous literature.}} Moreover, the black hole information paradox is brought up again and discussed.
\end{abstract}
\vfill {\footnotesize ~\\ xinyang\_wang@foxmail.com \\ 201621140020@mail.bnu.edu.cn \\ wbliu@bnu.edu.cn}
\newpage

\section{Introduction}
{\color{red}{In general relativity, it is a subtle issue to define the interior volume of a black hole which has a special physical significance, and it is related to the issue of choosing a particular spatial hypersurface for the volume. Parikh \cite{Parikh:2005qs} suggests that a reasonable definition of interior volume inside a black hole should be a slicing-invariant volume, which has been discussed by other authors \cite{Grumiller:2005zk, Ballik:2010rx, Ballik:2013uia, DiNunno:2008id, Finch:2012fq, Cvetic:2010jb, Gibbons:2012ac}. Recently, Christodoulou $et. al$ \cite{Christodoulou:2014yia} have proposed a definition of the interior volume, and introduced a special method to calculate the interior volume, which is similar to the method of calculating the geodesic equation of particles in the spacetime. Using this method, they found that the area of the spacelike hypersurface slice in a Schwarzschild black hole reaches its maximum value as $r= \frac{3}{2} m$, which is corresponding to the physical interior volume of a Schwarzschild black hole. Simultaneously, they also showed that the interior volume of an asymptotically flat Schwarzschild black hole with mass $m$, which is contained within the event horizon, is not static and grows like}}
\begin{equation}\label{vsch}
V \sim 3\sqrt{3} \pi m^2 v
\end{equation}
{\color{red}{when $v \gg m$,}} where $v$ is the advanced time. {\color{red}{As we all know,}} a classical Schwarzschild black hole remains static {\color{red}{when one has a view from outside, and has the constant area}} $16 \pi m^2$. {\color{red}{However, from inside,}} Eq. (\ref{vsch}) shows that {\color{red}{the interior volume of a classical Schwarzschild}} black hole grows with time. This result has been extended to the {\color{red}{RN}} \cite{Ong:2015dja} and the Kerr black holes \cite{Bengtsson:2015zda}. For the {\color{red}{RN}} black hole, since its geometry is spherical {\color{red}{which is similar to the Schwarzschild black hole, so that the formula of interior volume is also remarkably analogue to the Schwarzschild black hole, }}which means that the interior volume of {\color{red}{the RN}} black hole also increases linearly with the advanced time $v$. {\color{red}{However,}} for a Kerr black hole, {\color{red}{the geometry}} is much more complicated, {\color{red}{so that it is difficult to calculate its interior volume.}} To simplify this calculation, {\color{red}{Bengtsson $et. al$ \cite{Bengtsson:2015zda} directly}} chose an arbitrary hypersurface at constant $r$ inside a Kerr black hole {\color{red}{to calculate}} the volume of the hypersurface. {\color{red}{Although they used a different method, their results still suggest that when $r$ takes the special value $r_s$, the area of the hypersurface takes the maximal value, and corresponds to the interior volume.}}

Surprisingly, the interior volume of a black hole increases linearly with the advanced time $v$, {\color{red}{which}} can be a candidate to resolve the information paradox problem. Since the volume increases with time, a black hole {\color{red}{can}} have a large amount of volume to hide this remarkably huge information at the end of the evaporation process. {\color{red}{In this case, such a large volume may contain many models of quantum fields and the entropy of these models may relate to the Bekenstein-Hawking entropy. Naturally, one may ask that does the information hide in these fields and relate to their entropy? Therefore, it is necessary to investigate the entropy of the hidden modes of a field inside the black hole \cite{Kleban:2004rx, Gwak:2017kkt, Erdmenger:2017pfh}.}}  For a Schwarzschild black hole, the entropy has been studied in Ref.\cite{Zhang:2015gda}. The points which were taken {\color{red}{in Ref.\cite{Zhang:2015gda}}} are objectively described as follows. (i) The Klein-Gordon equation {\color{red}{is expanded}} under {\color{red}{a time-depending background. Using the WKB approximation,}} the scalar field can be written as $ \Phi = exp[-iET]exp[iI(\lambda, \theta, \phi)] $, where $E$ {\color{red}{denotes}} the energy of the scalar modes.  {\color{red}{(ii)}} Although the interior volume is dynamic, the statistical properties of scalar field modes in it can still be studied. The number of quantum states with energy less than $E$ is obtained by {\color{red}{integrating the quantum states in the phase space.}} The free energy and the entropy of the interior volume can be derived from the number of quantum states by using the standard statistical method. (iii) Hawking radiation is a thermodynamic process and the Stefan-Boltzmann law fits {\color{red}{well}} in it. Finally, the result verified that the entropy associated with the interior volume is proportional to the area of the horizon of the black hole.

In this paper, we propose a general method to calculate {\color{red}{the quantum states inside a black hole, which can be smoothly applied to the more general black holes. After introducing the two important assumptions as the blackbody radiation assumption and the quasi-static process assumption, the entropy of the scalar field inside a Kerr black hole has been calculated by using the differential form. Then, we compare the variation of the entropy of the scalar field with the variation of the Bekenstein-Hawking entropy and find that they are also proportional to each other in a Kerr black hole except the late stage of black hole evaporation. }}Subsequently, {\color{red}{the entropy of scalar field inside a Schwarzschild black hole}} is recalculated by using the differential form. The result shows that the proportionality coefficient between the entropy and the Bekenstein-Hawking entropy is half of that given in the previous literature. And we verify that if the angular momentum degenerates to zero, {\color{red}{the proportionality coefficient}} in the Kerr's case can {\color{red}{exactly go back to the result obtained by using the differential form in a Schwarzschild black hole.}} In the end, the black hole information paradox will be discussed in the view of the number of quantum states inside the black hole.

The organization of the paper is as follows. In the next section, {\color{red}{we review the definition of interior volume inside a black hole and choose the maximal hypersurface at constant $r$ which corresponds to the interior volume of a Kerr black hole.}} In section 3, we calculate the entropy of a certain region using the standard statistical method and {\color{red}{discuss the relationship between the entropy of the scalar field and the Bekenstein-Hawking entropy.}} In section 4, {\color{red}{using differential form,}} we derive the proportionality relationship between two kinds of entropy in a Schwarzschild black hole, and compare it with the Kerr's case when the angular momentum degenerates to zero. In section 5, we discuss the results and try to solve some problems about the black hole information paradox. By the way, some conclusions are given.

\section{The Interior Volume of a Kerr Black Hole}
{\color{red}{Before calculating the interior volume of a Kerr black hole, we first review the definition of the interior volume inside a spherically symmetric black hole proposed by Christodoulou $et.\, al$. In flat spacetime, the volume enclosed by the two-dimensional sphere $S$ is expressed as $V=\frac{4}{3} \pi R^3$, where $R$ is the radius of the two-sphere. It means that the volume inside a two-sphere $S$ is a three-dimensional spacelike hypersurface $\Sigma$ bounded by $S$. Although there are a lot of spacelike hypersurfaces bounded by $S$ in the spacetime, a special hypersurface exists whose volume is exactly the one that the two-sphere surrounds. Such a special hypersurface can be given in two equivalent manners:\\

(a) $\Sigma$ lies on the same simultaneity surface as $S$; \\

(b) $\Sigma$ is the largest spherically symmetric surface bounded by $S$. \\

\noindent
However, (a) cannot generalize the conception of volume inside the two-sphere to the curved spacetime, since the the simultaneity surface has no special significance in general. In contrast, (b) can be generalized to the curved case immediately. Therefore, if we want to obtain the interior volume of a spherically symmetric black hole, we just need to find the maximal spacelike hypersurface which is bounded by the two-sphere $S$. The Ref. \cite{Christodoulou:2014yia} has shown that the maximal hypersurface is obtained at $r=\frac{3}{2}m$, and corresponds to the interior volume inside a Schwarzschild black hole when $v \gg m$. In other words, the maximal hypersurface at constant $r$ is the interior volume of a Schwarzschild black hole.

Recently, the definition of interior volume inside a black hole has been extended from spherically symmetric black hole to axial symmetric cases \cite{Bengtsson:2015zda}. In the following, we will review how to calculate the interior volume of a Kerr black hole. The metric of a Kerr black hole in the Eddington-Finkelstein coordinates is \cite{Kerr:1963ud}}}
\begin{equation}\begin{split}
ds^2 = &-\frac{\Delta-a^2 sin^2\theta}{\rho^2} dv^2 + 2 dv \, dr + \rho^2 d\theta^2 + \frac{Asin^2\theta}{\rho^2} d\phi^2 \\
&- 2 a \, sin^2 \theta \, dr \, d\phi - \frac{4 a m r}{\rho^2} sin^2 \theta \, dv \, d\phi,
\end{split}\end{equation}
where
\begin{equation}
\begin{split}
\Delta &\equiv r^2-2mr+a^2,\\
\rho^2 &\equiv r^2 + a^2 cos^2\theta, \\
A &\equiv (r^2 + a^2)^2 - a^2 \Delta sin^2\theta,
\end{split}
\end{equation}
where $a=\frac{J}{m}$ is the angular momentum of unit mass of the black hole. The interior and exterior event horizons are $r_\pm = m \pm \sqrt{m^2 - a^2}$. For any one of the radially ingoing null geodesics, the advanced time $v$ is a constant and the coordinate $r$ {\color{red}{is physically interpreted}} as an affine parameter along the geodesic. The interior and exterior horizons of a Kerr black hole are not spherically symmetric, {\color{red}{which has some certain effects on the interior of the black hole, of which}} the primary one is that the singularity is no longer a point. {\color{red}{However, because the interior volume is far from the singularity when $v \gg m$, the special properties of a Kerr black hole do not affect the calculation of the interior volume. Therefore, according to the result of the Schwarzschild black hole, the interior volume of a Kerr black hole between two event horizons can be calculated.}} If we want to calculate the interior volume of a Kerr black hole, {\color{red}{we must obtain the induced metric on the spacelike hypersurface at constant $r$ inside a Kerr black hole.}}

An arbitrary vector on the hypersurface at constant $r$ can be decomposed into two parts: one is normal to the hypersurface, the other is the tangent vector of the hypersurface. For example, an arbitrary vector can be expressed as
 \begin{equation}
t^a = N n^a + N^a,
\end{equation}
where $N$ and $N^a$ are the lapse-function and the shift vector respectively. The covector on the hypersurface is defined as
\begin{equation}\label{na}
n_a = -N \nabla_a r,
\end{equation}
where $\nabla_a r$ is the normal covector and its equivalent expression is $(dr)_a$. {\color{red}{Using the normalization condition $n_a n^a =-1$, we obtain the relation
\begin{equation}
N =\sqrt{\frac{-1}{g^{rr}}}.
\end{equation}
According to the condition $h=N^{-2}(-g)$, where $h$ is the determinant of induced metric on the hyperfurface and $g$ is the determinant of metric in spacetime, we obtain the volume for an arbitrary hypersurface at constant $r$,}}
\begin{equation}\label{Volume}
\begin{aligned}
V_\Sigma =& \int \sqrt{h} dv \wedge d\theta \wedge d\phi \\
=&2 \pi v \sqrt{- \Delta} \left(\sqrt{r^2 + a^2} + \frac{r^2}{2a} \ln \frac{\sqrt{r^2 + a^2} + a}{\sqrt{r^2 + a^2} - a} \right).
\end{aligned}
\end{equation}
{\color{red}{If we want to obtain the interior volume of a Kerr black hole, the maximal hypersurface at constant $r$ must be found. It means that the condition $\frac{\partial V_{\Sigma}}{\partial r}=0$ should be satisfied. The solution of this condition is $r=r_{s}$ which corresponds to the maximal hypersurface at constant $r$. Therefore, substituting $r=r_{s}$ into the Eq. (\ref{Volume}), the interior volume of a Kerr black hole is obtained.}}
\section{Entropy in the Volume of a Kerr Black Hole}
Although a classical black hole remains stationary and it always has the same area of event horizon from the exterior point of view, its {\color{red}{interior volume}} grows with advanced time $v$. {\color{red}{The property of the interior volume is different from that of the volume of the black hole on which an outside observation is taken. The special character of the interior volume may influence the statistical quantities of the quantum field inside the black hole and it may propose a solution to the information paradox of black hole.}} Hence, it is significant to investigate {\color{red}{how the special character of interior volume influences the statistical quantities of the quantum fields.}} In this paper, we only involve the massless scalar field {\color{red}{inside the black hole. In order to get the statistical quantities in the interior volume, we have to construct the coordinates on the maximal hypersurface at constant $r$ firstly, then calculate the number of quantum states and other statistical quantities in the phase space using the method of equilibrium statistics.}}

{\color{red}{In general, we can choose the coordinate $\{x^1, x^2, x^3 \}$ on the maximal hypersurface at constant $r$, on which the induced metric is
\begin{equation}
d\hat{s}^2 = h_{ij} dx^i dx^j.
\end{equation}
For any point $p$ on the hypersurface, there must exist a Gaussian normal coordinate system $\{T, x_p^i \}, i=1, 2, 3$, defined by a family of geodesics, where $T$ is the affine parameter of the geodesic, and $x_p^i$ denotes the spatial coordinate of point $p$. Thus, the line element can be expressed as
\begin{equation}
ds^2 = -dT^2 + h_{ij} dx^i dx^j.
\end{equation}
Now, we consider the massless scalar field $\Phi$ in this spacetime. Actually, the line element is equivalent to the form $ds^2 = -dT^2 + h_{ij}(T) dx^i dx^j$,}} which means {\color{red}{that the hypersurface at constant $r$ in the the black hole is dynamical for the defined time $T$. However, the method of equilibrium statistics to calculate the thermodynamic quantities of the system can be used only if the background is static, so that we cannot use the method to investigate the statistical property of the scalar field in such a dynamical situation. Fortunately, according to Ref. \cite{Christodoulou:2016tuu}, when $K=0$, the corresponding hypersurface is called the maximal hypersurface, where $K$ is the trace of the extrinsic curvature of the hypersurface. From $K=0$, we can derive the relation as }}
\begin{equation}
{\color{red}{\mathcal{L}_n h=0.}}
\end{equation}
{\color{red}{According to this relation, we can infer $\mathcal{L}_n \hat{\epsilon}=0$, where $\hat{\epsilon}$ is the induced volume element of the hypersurface. }}That is to say, the maximal hypersurface at constant $r$ does not change with the proper time, then the properties are calculated on the hypersurface at $T=constant$ which corresponds to the  interior volume of the black hole. {\color{red}{So that our statistical calculation is not affected by the nonstatic character of the metric.}} Next, we will use the common method in the curved spacetime to discuss the motion of the scalar field in the interior of the black hole.

Using the equation of motion of the massless scalar field, we obtain
\begin{equation}
P^\mu P_\mu = g^{\mu \nu} P_\mu P_\nu = g^{00}E^2 + h^{ij} P_i P_j=-E^2 +  h^{ij} P_i P_j=0,
\end{equation}
{\color{red}{in which $g^{\mu \nu}$ is the inverse metric of spacetime on the hypersurface at constant $r$ and $h^{ij}$ is the inverse induced metric which is non-diagonal in general case.}} Since $h^{ij} P_i P_j$ is a quadratic form, we can always find a similarity transformation which can diagonalize the induced metric. Completing the transformation, we can obtain
\begin{equation}
E^2 - \lambda^1 {P'_1}^2 -\lambda^2 {P'_2}^2 - \lambda^3 {P'_3}^2 = 0,
\end{equation}
{\color{red}{where $\lambda^1, \lambda^2$ and $\lambda^3$ are the diagonal elements of the diagonalizable induced metric respectively, and $P'_1, P'_2$ and $P'_3$ are the elements of the eigenstate of the diagonalizable induced metric. }}
Thus the number of quantum states with energy less than $E$ can be obtained as
{\color{red}{\begin{equation}
\begin{aligned}\label{g(E)}
 g(E)&=\frac{1}{(2 \pi)^3} \int dx_1 dx_2 dx_3 dP_1 dP_2 dP_3\\
&=\frac{1}{(2 \pi)^3} \int dx_1 dx_2 dx_3 dP'_2 dP'_3 \sqrt{\frac{1}{\lambda^1}} \sqrt{E^2 - \lambda^2 {P'_2}^2 - \lambda^3 {P'_3}^2}\\
&=\frac{E^3}{(2 \pi)^3} \int dx_1 dx_2 dx_3 \frac{2 \pi}{3} \frac{1}{\sqrt{\lambda^1 \lambda^2 \lambda^3}}\\
&=\frac{E^3}{12 \pi^2} V_\Sigma,
\end{aligned}
\end{equation}}}
where the relation $P'_1  = \sqrt{\frac{1}{\lambda^1}} \sqrt{E^2 - \lambda^2 {P'_2}^2 - \lambda^3 {P'_3}^2}$ is used in the second line, the integral formula $\iint \sqrt{1 - \frac{x^2}{a^2} - \frac{y^2}{b^2}} dx dy = \frac{2 \pi}{3} a b$ is used in the third line. {\color{red}{We can see that the number of  quantum states is proportional to the interior volume of a Kerr black hole. This result appears to be very similar to the result in the flat spacetime. However, their physical significance is very different. In the flat spacetime, because the volume  bounded by a closed spacelike hypersurface is a constant, then the number of quantum states which is proportional to the volume is a constant. However, the volume inside the black hole is no longer a constant and it can increase with the advanced time $v$, then the number of quantum states of the scalar field inside the black hole can also increase with the advanced time. This property of the quantum states is the special characteristic of the curved spacetime, and it will influence the results in the following.

Subsequently, we continue to calculate the free energy of the scalar field at some inverse temperature $\beta$ as
\begin{equation}
\begin{aligned}
 F(\beta)&=\frac{1}{\beta} \int dg(E) \ln(1 - e^{-\beta E})\\
&=\frac{V_\Sigma}{12 \pi^2} \int \frac{E^3 dE}{e^{\beta E}-1}\\
&=- \frac{\pi^2 V_\Sigma}{180 \beta^4}.
\end{aligned}
\end{equation}
Furthermore, the entropy is obtained as
\begin{equation}\label{entropy}
 S_\Sigma=\beta^2 \frac{\partial F}{\partial \beta}=\frac{\pi^2 V_\Sigma}{45 \beta^3},
\end{equation}
where $\beta$ is the inverse of the scalar filed's temperature. This entropy is similar to that in the flat phase space. However, it also increases with the advanced time $v$ because it is proportional to the interior volume of a black hole. }}

The above calculation leads to the conclusion that {\color{red}{the number of quantum states and the entropy of the scalar field inside the black hole are independent on the specific form of metrics. So that the method of calculating these properties can be applicable to any other black holes. }}Therefore, this method has its own merit in calculating the entropy of the black hole in general cases.

{\color{red}{Next, the two important assumptions will be taken into account\cite{Zhang:2015gda, Majhi:2017tab, Zhang:2017aqf}, which can be summarized as follows:
\\

\hangafter 1
\hangindent 1.5em
\noindent
(a) The blackbody radiation assumption. The Hawking radiation of a black hole can be
seen as black body radiation. Then, the radiation temperature at infinity can be considered
as the temperature of event horizon. \\ \\

\hangafter 1
\hangindent 1.5em
\noindent
(b) The quasi-static process assumption $\frac{dm}{dv} \ll 1$, which means that the evaporation process is slow enough. Although the Hawking temperature is changing slowly, because of this condition the thermal equilibrium between the scalar field inside the black hole and the event horizon is preserved in this adiabatic process. \\

\noindent
According to the assumption (a), the temperature of event horizon can be regarded as the Hawking temperature. And then based on the assumption (b), the temperature of both the scalar field and the horizon can be regarded as equal when they can establish a thermal equilibrium in an infinitesimal process. Therefore, the temperature of the scalar field inside the black hole can be considered as the Hawking temperature,  and $\beta$, the inverse of the temperature, can be expressed as
\begin{equation}\label{beta}
\beta = \frac{1}{T} = \frac{2 \pi [(m(v) + \sqrt{m(v)^2 - a(v)^2})^2 + a(v)^2]}{\sqrt{m(v)^2 - a(v)^2}}.
\end{equation}}}

Now we {\color{red}{consider}} the interior volume of a Kerr black hole {\color{red}{with the Hawking radiation. According to the assumption (a),}} the lost mass rate of a Kerr black hole can be given by the Stefan-Boltzmann law \cite{Montvay:1981jj, QFTCS}
\begin{equation}
\frac{dm}{dv} = -\frac{1}{\gamma} T^4 A,  \qquad \gamma > 0,
\end{equation}
where $\gamma$ is a positive constant that depends on the number of quantized matter fields coupling with gravity, and its value does not influence the following discussion. In Ref. \cite{Ong:2015dja}, it has been discussed that the large volume remains until the final stage of black hole evaporation. At this point, the radiation can last. {\color{red}{Thus, for a black hole with mass $m(v)$ and angular momentum $a(v)$, we have
\begin{equation}\label{dv}
dv =  -\gamma  \frac{4 \pi^3 \left[2 m(v)^2 + 2m(v) \sqrt{m(v)^2 - a(v)^2}\right]^3}{\left[m(v)^2 - a(v)^2 \right]^2} dm,
\end{equation}}}
in which, due to the Hawking radiation, the mass is not constant and changes with the advanced time $v$. 

\begin{figure}[H]
\centering
\includegraphics[width=0.8\textwidth]{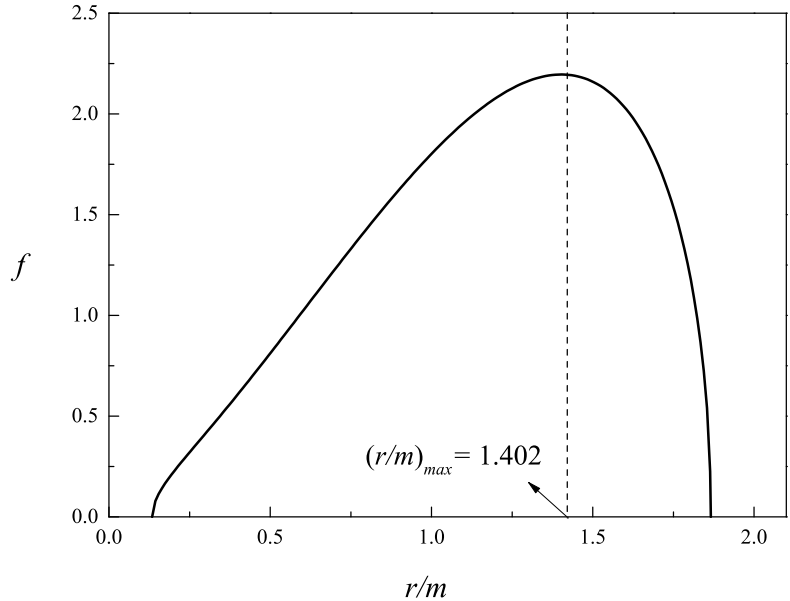}
\caption{Plot of $\frac{r}{m}$ versus $f \left(\frac{r}{m}\right)$. The function reaches its maximal value at $ \left( \frac{r}{m} \right)_{s} = 1.402$ when $\frac{a}{m}=0.5$.}\label{fig1}
\end{figure}

Subsequently, the interior volume of a Kerr black hole is calculated. Eq. (\ref{Volume}) can be written as
\begin{equation}\label{vsfr}
V_\Sigma = 2 \pi f\left(\frac{r}{m}, \frac{a}{m} \right) m^2 v,
\end{equation}
where
\begin{equation}\label{fra}
\begin{split}
f \left(\frac{r}{m}, \frac{a}{m} \right) = &\sqrt{2 \frac{r}{m} - \left(\frac{r}{m} \right)^2 - \left (\frac{a}{m} \right)^2 } \\
&\left[ \sqrt{\left(\frac{r}{m}\right)^2 + \left(\frac{a}{m}\right)^2} + \frac{m}{2 a} \left(\frac{r}{m}\right)^2 \ln \left(\frac{\sqrt{(\frac{r}{m})^2 + \left (\frac{a}{m} \right)^2}+ \frac{a}{m}}{\sqrt{(\frac{r}{m})^2 + \left (\frac{a}{m} \right)^2} -\frac{a}{m}} \right) \right].
\end{split}
\end{equation}
According to the previous statements, the hypersurface is at $r=r_s$, which corresponds to the {\color{red}{interior volume}} of a black hole. Therefore, {\color{red}{for Eq. (\ref{vsfr}),}} we just need to find out the special value of the $f \left( \frac{r}{m},\frac{a}{m} \right)$ as $\frac{r}{m} = \left(\frac{r}{m} \right)_{s}$, which corresponds to {\color{red}{the interior volume}} of a Kerr black hole. According to Eq. (\ref{fra}), we find the relationship between $\frac{r}{m}$ and $f \left(\frac{r}{m},\frac{a}{m} \right)$, as shown in Figure \ref{fig1}.
\noindent
From Figure \ref{fig1}, we can see that $f \left( \frac{r}{m},  \frac{a}{m}\right)$ indeed reaches its maximal value at $\left( \frac{r}{m} \right)_{s} = 1.402$, which corresponds to the interior volume of the Kerr black hole when $\frac{a}{m}=0.5$.

{\color{red}{At the end of this section, when we calculate the relationship between the entropy of scalar field and the Bekenstein-Hawking entropy, two difficulties emerge. Firstly, when we investigate the variation of entropy in an interval, the integral of the Eq. (\ref{dv}) is unavoidable. This integral is very complicated, so we cannot obtain the analytic solution. More importantly, the assumption (b) is no longer valid when we consider the evaporation in an interval, so the temperature of the scalar field inside the black hole cannot be considered as the Hawking temperature and the method of equilibrium statistics cannot be used. Secondly, when we calculate the entropy of scalar filed using the method of  equilibrium statistics, the temperature of the scalar field varies with the mass of black hole in the evaporation process, so that we cannot calculate the entropy of scalar field using the method of equilibrium statistics. 

However, if we consider an infinitesimal process to investigate the variation of the entropy of the scalar field, we can naturally avoid the two difficulties above. According to the assumptions (a) and (b), the temperature of the scalar field inside the black hole can be treated as the Hawking temperature. Meanwhile, in an infinitesimal process, the relationship between the scalar field and the horizon can be regarded as thermal equilibrium, so the temperature of the black hole can be regarded as a constant which is exactly the value at the beginning of the infinitesimal process. Since the Hawking temperature is a function of the mass, the mass can also be considered a constant in the infinitesimal process. Hence, the variation of both interior volume and the entropy only depends on the variation of advanced time $v$, meanwhile, the variation of the mass can be ignored. For this reason, the method of equilibrium statistics in the infinitesimal process can be used to discuss the variation of the scalar field's entropy. From the above, we can use the differential form instead of the integral form to investigate the entropy of the scalar field and the Bekenstein-Hawking entropy when the Hawking radiation is considered. In the following, we set the constant mass and angular momentum at the beginning of an infinitesimal process as $m_0(v)$ and $a_0(v)$ respectively.

According to the above statement, we differentiate two sides of the Eq.(\ref{vsfr}) and substitute both the maximal value of $f \left( \frac{r}{m}, \frac{a}{m} \right)$ and Eq.(\ref{dv}) into it, then the differential form of the interior volume is expressed as
\begin{equation}\label{vga}
\dot{V}_\Sigma = - 64 \pi^4 \gamma f_{max} \left(\frac{a_0(v)}{m_0(v)} \right) \frac{\left[1 + \sqrt{1 - \left(\frac{a_0(v)}{m_0(v)} \right)^2} \right]^3}{\left[1 - \left(\frac{a_0(v)}{m_0(v)} \right)^2 \right]^2} m_0^4(v) \,  \dot{m}(v),
\end{equation}
where $\dot{V}_\Sigma$ and $\dot{m}(v)$ represent $\frac{d V_\Sigma}{dv}$ and $\frac{dm(v)}{dv}$ respectively, $f_{max} \left(\frac{a_0(v)}{m_0(v)} \right)$ is the maximal value when $\left( \frac{r_0(v)}{m_0(v)} \right) = \left( \frac{r_0(v)}{m_0(v)} \right)_{s}$ and $\dot{m}(v) < 0$. }}Substituting both Eq.(\ref{beta}) and Eq.(\ref{vga}) into Eq.(\ref{entropy}), {\color{red}{we can obtain the differential form of the entropy in the infinitesimal process
\begin{equation}\label{fentropy}
\dot{S}_{\Sigma} = - \frac{8 \pi^3}{45} \gamma \frac{f_{max} \left(\frac{a_0(v)}{m_0(v)} \right) \left(1 + \sqrt{1 - \left(\frac{a_0(v)}{m_0(v)} \right)^2} \right)^3}{\sqrt{1-\left(\frac{a_0(v)}{m_0(v)} \right)^2} \left[\left(1+\sqrt{1-\left(\frac{a_0(v)}{m_0(v)} \right)^2} \right)^2 + \left(\frac{a_0(v)}{m_0(v)} \right)^2 \right]^3} m_0(v) \, \dot{m}(v).
\end{equation}}}

\begin{figure}[H]
\centering
\includegraphics[width=0.8\textwidth]{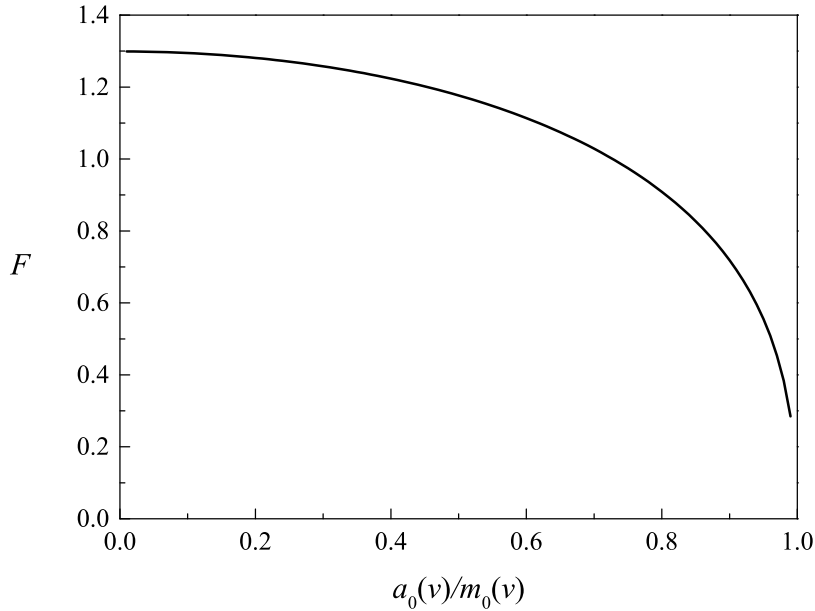}
\caption{Plot of $\frac{a_0(v)}{m_0(v)}$ versus $F\left(\frac{a_0(v)}{m_0(v)}\right)$. This figure shows that $F\left(\frac{a_0(v)}{m_0(v)}\right)$ can be approximated as a constant at the beginning of the evaporation of the black hole, but the value of $F\left(\frac{a_0(v)}{m_0(v)}\right)$ decreases rapidly with the change of mass at the end of black hole evaporation.}\label{Fig2}
\end{figure}

Finally, we calculate the variation of the Bekenstein-Hawking entropy and compare it with the variation of the entropy of the scalar field in the infinitesimal process. The Bekenstein-Hawking entropy is defined as \cite{Hawking:1974sw, Bekenstein:1972tm, Bekenstein:1973ur}
\begin{equation}
 S_{BH} = \frac{A}{4},
\end{equation}
where $A = 4 \pi (r_+^2 + a^2)$ is the area of event horizon of a Kerr black hole. {\color{red}{Using the differential form, the variation of the Bekenstein-Hawking entropy in this process can be expressed as 
\begin{equation}\label{dsbh}
\dot{S}_{BH} = 4 \pi \frac{1+\sqrt{1-\left(\frac{a_0(v)}{m_0(v)} \right)^2}}{\sqrt{1- \left(\frac{a_0(v)}{m_0(v)} \right)^2}} m_0(v) \, \dot{m}(v),
\end{equation}
in which $\dot{m}(v)$ is also negative.}} Substituting Eq.(\ref{dsbh}) into Eq.(\ref{fentropy}), we {\color{red}{obtain}} the differential relationship between the entropy of interior volume and the Bekenstein-Hawking entropy 
{\color{red}{\begin{equation}\label{two types entropy}
\dot{S}_\Sigma = -\frac{\pi^2}{180} \gamma F \left(\frac{a_0(v)}{m_0(v)} \right)   \dot{S}_{BH},
\end{equation}
where
\begin{equation}\label{F}
F \left(\frac{a_0(v)}{m_0(v)} \right) = f_{max} \left(\frac{a_0(v)}{m_0(v)} \right) \left[1- \sqrt{1-\left(\frac{a_0(v)}{m_0(v)} \right)^2} \right] \left(\frac{a_0(v)}{m_0(v)} \right)^{-2}.
\end{equation}}}
According to Eq.(\ref{F}), we have the relationship between the $\frac{a_0(v)}{m_0(v)}$ and $F \left(\frac{a_0(v)}{m_0(v)}\right)$, as shown in the Figure \ref{Fig2}.
{\color{red}{Figure \ref{Fig2} shows that the function $F\left(\frac{a_0(v)}{m_0(v)}\right)$ can be approximated as a constant in the early stage of the evaporation, which means that the variation of entropy from quantum theory in the interior volume is proportional to the variation of the Bekenstein-Hawking entropy. It means that the expansion of interior volume is proportional to the shrink of event horizon for a Kerr black hole.This is an intriguing result. However, the function $F\left(\frac{a_0(v)}{m_0(v)}\right)$ with the change of mass reduces greatly at the end of the evaporation of the black hole, which means the variation of the two kinds of entropy violate the  proportional relationship. The value of  $F\left(\frac{a_0(v)}{m_0(v)}\right)$ is not suitable for the late stage of the black hole evaporation. Since the process can no longer be considered as a very slow process in the late stage of black hole evaporation, the assumption (b) is not satisfied. Based upon the method of calculation, if we want to calculate the entropy of the scalar field inside the black hole at the end of the Hawking radiation, the variation of the mass of the black hole cannot be ignored in the infinitesimal process.  In other words, when we differentiate two sides of the Eq. (\ref{vsfr}), the mass of the black hole cannot be considered as a constant. Since the area of the event horizon becomes very small, the effect of quantum gravity cannot be ignored at the end of the black hole evaporation as well. For this reason, the calculation of the scalar filed's entropy inside the black hole at the end of the evaporation process becomes very difficult. Therefore, we will discuss the situation of the late stage of the black hole evaporation in subsequent research work.}}

\section{Entropy in the Volume of a Schwarzschild Black Hole}
{\color{red}{In this section, we return back to the Schwarzschild black hole.}} The relationship between the entropy of scalar field inside the black hole and the Bekenstein-Hawking entropy has been studied in Ref. \cite{Zhang:2015gda}. However, the method used to calculate the entropy of scalar field inside the black hole is flawed in the literature, because it chose an unreasonable temperature to calculate the entropy of scalar field. {\color{red}{The primary reason beneath this issue is that when the Hawking radiation is considered, the black hole's mass varies with time, which leads to a change in the temperature of the black hole as time goes on.}} Therefore, the method of equilibrium statistics cannot be applied in this situation. {\color{red}{To solve this problem, according to the assumptions (a) and (b), we use the differential form which is similar to the Kerr case to calculate the relationship between the variation of the entropy of the scalar field inside the black hole and the variation of the Bekenstein-Hawking entropy in the infinitesimal process. The proportionality coefficient between the two kinds of entropy can be obtained more reasonably by using the differential form, and the result obtained in the previous literature are corrected. }}

We also start from the interior volume of a Schwarzschild black hole and it can be expressed as\cite{Christodoulou:2014yia}
\begin{equation}\label{SV}
V_{\Sigma} \sim 3\sqrt{3} \pi m^2 v.
\end{equation}
{\color{red}{By adopting the method of equilibrium statistics, we can obtain the entropy of scalar field in the volume.}} The expression of entropy is the same as Eq. (\ref{entropy}). {\color{red}{According to the assumption (a),}} a Schwarzschild black hole evaporation can also be seen as black body radiation. Therefore, we can use the Stefan-Boltzmann law
\begin{equation}\label{Sdmdv}
\frac{dm}{dv} = -\frac{1}{\gamma} T^4 A
\end{equation}
to calculate the change of mass with the advanced time $v$, where $A=16 \pi m^2(v)$ is the area of event horizon. {\color{red}{According to the assumptions (a) and (b), the temperature of the scalar field can be regarded as the Hawking temperature, which is expressed as
\begin{equation}\label{ST}
T = \frac{1}{8 \pi m(v)}.
\end{equation}}}
Substituting Eq. (\ref{ST}) into Eq. (\ref{Sdmdv}), we have
{\color{red}{\begin{equation}\label{Sdv}
dv=-2^8  \gamma \pi^3 m^2(v) \, dm.
\end{equation}}}
{\color{red}{According to the assumption (b), the Hawking radiation is very slow and the evaporation process can be regarded as a quasi-static process. Hence, the variation of the interior volume of the black hole only depends on the advanced time $v$. }}Differentiating two sides of the Eq. (\ref{SV}) and substituting Eq. (\ref{Sdv}) into it, we can obtain the interior volume in an infinitesimal process which can be expressed as
{\color{red}{\begin{equation}\label{SdV}
\dot{V}_{\Sigma} = - 2^8 \times 3\sqrt{3} \pi^4 \gamma m_0^4(v) \,  \dot{m}(v).
\end{equation}}}
Substituting Eq. (\ref{SdV}) into Eq. (\ref{entropy}), we can obtain the differential form of entropy inside the black hole
{\color{red}{\begin{equation}\label{SdSS}
\dot{S}_{\Sigma} = - \frac{\sqrt{3} \pi^3 \gamma}{30}  m_0(v) \, \dot{m}(v).
\end{equation}}}
From the definition of the Bekenstein-Hawking entropy, we can directly derive its differential form as
{\color{red}{\begin{equation}\label{SdSB}
\dot{S}_{BH} = 8 \pi m_0(v)\, \dot{m}(v).
\end{equation}}}
Combining Eq. (\ref{SdSB}) with Eq. (\ref{SdSS}), we find the differential relationship between the entropy of interior volume and the Bekenstein-Hawking entropy, which can be expressed as
{\color{red}{\begin{equation}\label{Snum}
\dot{S}_{\Sigma}=-\frac{\sqrt{3} \pi^2 \gamma }{240} \dot{S}_{BH}.
\end{equation}}}

\begin{figure}[H]
\centering
\includegraphics[width=0.8\textwidth]{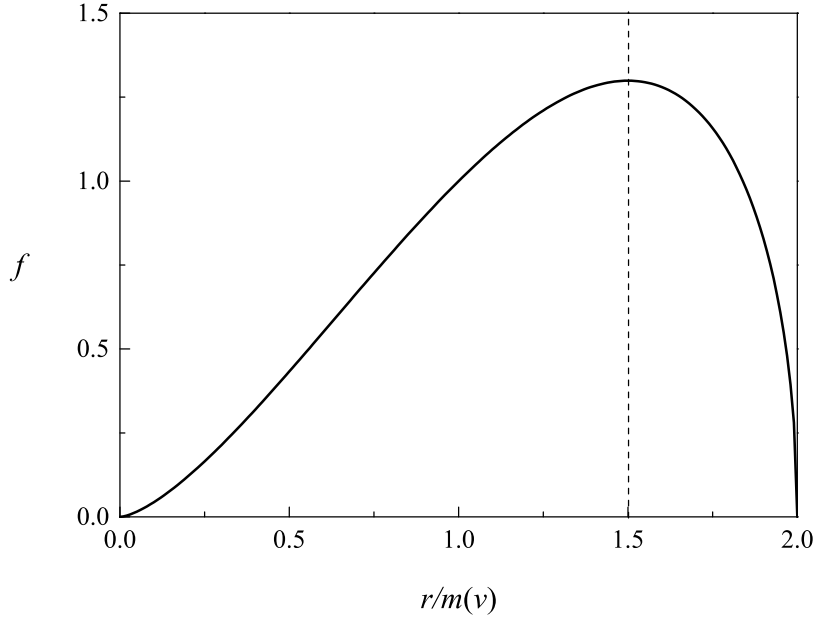}
\caption{Plot of $\frac{r}{m(v)}$ versus $f\left(\frac{r}{m(v)}\right)$. This figure shows that $f\left(\frac{r}{m(v)}\right)$ has maximal value at $\frac{r}{m(v)}=\frac{3}{2}$.}\label{fig3}
\end{figure}

Finally, if we set $a = 0$, the results in a Kerr black hole are expected to degenerate to the results in a Schwarzschild black hole. In this approach, {\color{red}{we verify}} that it can obtain the reasonable result by using the differential form to study the Hawking radiation in these two kinds of black hole. Therefore, the limit of Eq. (\ref{F}) can be expressed as
\begin{eqnarray}
\begin{split}
&\underset{\frac{a}{m} \to 0}{\text{lim}} f \left(\frac{r}{m(v)},\frac{a(v)}{m(v)} \right) \left[1- \sqrt{1-\left(\frac{a(v)}{m(v)} \right)^2} \right] \left(\frac{a(v)}{m(v)} \right)^{-2}\\
=& \frac{r}{m(v)} \sqrt{2\frac{r}{m(v)}-\left(\frac{r}{m(v)} \right)^2}\\
=&f\left(\frac{r}{m(v)} \right),
\end{split}
\end{eqnarray}
{\color{red}{where $\frac{r}{m(v)}$ must take special value, because it corresponds to the interior volume of the black hole.}} The relationship between the $\frac{r}{m(v)}$ and the value of the limit can be shown in Figure \ref{fig3}.
{\color{red}{Figure \ref{fig3} illustrates that the function $f_{max}\left(\frac{r}{m(v)} \right)$ has maximal value at $\frac{r}{m(v)}=\frac{3}{2}$, and this value is exactly the same as the Schwarzschild case.}} Substituting the maximal value of  $f\left(\frac{r}{m(v)}\right)$ into the Eq. (\ref{two types entropy}), we obtain the proportional relation between two types of entropy
{\color{red}{\begin{equation}\label{Knum}
\dot{S}_{\Sigma}=-\frac{\sqrt{3} \pi^2 \gamma}{240}  \dot{S}_{BH}.
\end{equation}}}
Comparing Eq. (\ref{Knum}) with Eq. (\ref{Snum}), we find that they are exactly the same. This means the proportional relation in the Kerr case can totally degenerate to the Schwarzschild case. This complete degeneration relationship reflects the interior volume of a Kerr black hole can be degenerated to that of a Schwarzschild black hole when $a=0$ {\color{red}{and it also reflects}} the fact that the proportional relationship between the variation of the two kinds of entropy for a Schwarzschild black hole is a special case for a Kerr black hole.

\section{Discussions and Conclusions}
{\color{red}{According to the two important assumptions, we have adopted the differential form for discussing the variation of entropy inside a Kerr black hole in an infinitesimal process, and have obtained the relationship between the variation of entropy of the scalar filed inside the black hole and the variation of the Bekenstein-Hawking entropy.}} The analysis and discussion of the result show that the proportionality coefficient of the two kinds of entropy is approximated as a constant {\color{red}{except the late stage of the evaporation process.}} In other words, the variation of entropy of the scalar field is proportional to the variation of the Bekenstein-Hawking entropy. {\color{red}{However, the two kinds of entropy do not satisfy this relation at the end of the evaporation, since the assumption (b) does not hold. }}

{\color{red}{We also have adopted the differential form to calculate the variation of scalar field entropy inside the Schwarzschild black hole when the Hawking radiation is considered, and have calculated the proportional relation between it and the variation of the Bekenstein-Hawking entropy. Meanwhile, setting $a=0$, we expect that the result of the Kerr case can completely degenerate to the Schwarzschild case to verify the results obtained.}} By comparing the two results, it reveals that, if the angular momentum degenerates to zero, the proportional relation between the variation of two types of entropy in the Kerr black hole is exactly the same as the result calculated by the differential form in the Schwarzschild black hole. This conclusion reflects that {\color{red}{with the Hawking radiation,}} if we use the differential form to calculate the proportional relation between the variation of the entropy of the scalar field inside the black hole and the variation of the Bekenstein-Hawking entropy, we {\color{red}{will}} obtain the reasonable results.

In the previous literature \cite{Zhang:2015gda}, it is unreasonable to use integral method to study the proportional relation between the two types of entropy. There are two reasons for the unreasonableness. {\color{red}{Firstly, the quasi-static assumption cannot be used in the interval. Secondly, the changing temperature with the advanced time $v$ is replaced by the initial temperature. Based on these two reasons,}} the unreasonable proportional relation can be obtained. However, if the differential form is used, the unreasonable situation can be naturally avoided. The proportional coefficient obtained by the differential method is half of that obtained by the integral method. {\color{red}{We corrected the result in the previous literature by improving the calculation method.}}

According to previous calculations, the interior volume of a Kerr black hole grows with the advanced time $v$ for $v \gg m$. If the massless scalar field is added inside the black hole, the number of the quantum states of the scalar field is proportional to the interior volume. While the interior volume increases with the advanced time $v$, so does the number of quantum states. It is really a clue to think about the black hole information paradox \cite{Rovelli:2014cta, Rovelli:1996ti, Rovelli:1996dv, Strominger:1996sh, York:1983zb, Zurek:1985gd, tHooft:1990fkf, Susskind:1993if, Frolov:1993ym, Carlip:1994gy, Cvetic:1995bj, Larsen:1995ss, Horowitz:1996fn, Strominger:1997eq, tHooft:2016fzb, Almheiri:2012rt}.

In the process of black hole evaporation, the mass loss of black hole is due to Hawking radiation, then the surface area of the event horizon decreases, which accounts for the decrease of the Bekenstein-Hawking entropy. However, since the entropy of the scalar field in the interior of black hole is proportional to the interior volume of the black hole, the entropy of the scalar filed also increases while the interior volume increases with the advanced time $v$. The relationship between these two types of entropy is expressed as Eq. (\ref{two types entropy}) and Eq. (\ref{Snum}).

It is often assumed that the maximal number of quantum states contained in a black hole surface is \cite{Page:1993wv, Marolf:2017jkr, Jacobson:2005kr}
\begin{equation}
N_{BH} = e^{S_{BH}},
\end{equation}
where $ S_{BH} = \frac{A}{4}$ is the Bekenstein-Hawking entropy and $A$ is the area of the event horizon. When the evaporation happens, the number of quantum states on the black hole surface gradually reduces due to the decrease of the Bekenstein-Hawking entropy. According to Eq.(\ref{g(E)}), the number of quantum states inside the black hole increases with the expansion of the interior volume, and the entropy of the scalar field in the interior volume also increases. When the black hole evaporation eventually stops, the number of quantum states inside the black hole is much more than the number of quantum states on the surface. Hence, the results in this paper are consistent with Ref. \cite{Rovelli:2017mzl}. According to Ref.\cite{Ong:2015dja}, the large volume inside the black hole should be able to contain all the information, even though the area of event horizon shrinks to a very small size. In the end, the black hole becomes "a remnant" \cite{Chen:2014jwq, Aharonov:1987tp}, and it has enough room to store the information. Based on this fact, we propose that the information loss claimed in the process of black hole evaporation is actually stored in the interior of the black hole as the form of quantum states of the scalar field.

{\color{red}{In this paper, the interior volume of a Kerr black hole has been investigated.}} Moreover, some explanation about why the statistical quantities can be calculated in the dynamical background {\color{red}{is given}}. Involving the massless scalar field inside the black hole, we {\color{red}{have}} proposed a more general method to calculate {\color{red}{the number of quantum states}} in the interior volume of a Kerr black hole, and found that the entropy of a Kerr black hole is deeply related to the Bekenstein-Hawking entropy. {\color{red}{Based on the two assumptions, the proportional relation between the two types of entropy has been calculated by using the differential form. It is found that they are proportional to each other except the late stage of the black hole evaporation. Moreover, the proportional relation in Schwarzschild black hole has been recalculated. It is found that the proportionality coefficient of the two types of entropy is half of the result obtained in the previous literature and the proportionality coefficient in a Kerr black hole can completely degenerate to the Schwarzschild case when the angular momentum degenerated to zero. It means that the reasonable result can be obtained by using the differential form and the result obtained in the previous literature is corrected.}} Moreover, we have explained the black hole information paradox from the number of quantum states inside the black hole. Early literature reveals that the number of quantum states on the black hole surface is bounded. However, our calculation have shown that the number of quantum states of the scalar field inside the black hole is much more than that on the surface. When Hawking radiation exists, the number of quantum states on the surface gradually decreases. On the contrary, the number of quantum states inside the black hole increases. Consequently, we propose that the information so-called lost in the process of black hole evaporation is stored in the black hole as the form of quantum states. Hence, the information has never been lost.

\section*{Acknowledgement}
This work is supported by the National Natural Science Foundation of China (Grant No. 11235003).

\end{document}